# Evaluation of Porosity for Gamma Irradiated Poly(ethylene oxide): A New Approach Using Microscopic Image Aided with Computer Programming

Subir K. Patla[1,4], Madhumita Mukhopadhyay*[2], Abhra Giri*[3], Ruma Ray[4],

and Sujata Tarafdar[1]

[1]Department of Physics,Jadavpur University, Condensed Matter Physics Research Centre, Kolkata – 700032, WB, India

[2] Department of Chemistry, Amity Institute of Applied Sciences (AIAS), Amity University, Kolkata – 700135, WB, India

[3]Department of Physics, L. J. D. College, Falta, Sahararhat, South 24 Parganas – 743504, WB, India

[4]Department of Physics,Gurudas College, Kolkata – 700054, WB, India

Correspondence to: madhubanerji@gmail.com (M. Mukhopadhyay)

**Tel:**+919433882877

**Fax:** +91 3324148917.

and

abhrag321@gmail.com (A. Giri)

**Tel:**+919748330712

E-mail Addresses: subirpatla@gmail.com, madhubanerji@gmail.com, abhrag321@gmail.com ,me.rumaray@gmail.com and sujata_tarafdar@hotmail.com.

**Abstract**

Porous polymer films of Poly(ethylene oxide)(PEO) are synthesized by solution casting technique using gamma irradiated: (a) PEO powder (S-series) &(b) PEO - methanol solution (L-series). Pore phase is though a defect but form instantaneously during preparation. Optimization of the pore content and distribution enables tailoring of associated properties for multifaceted applications. The experimental pore-size and distribution of the PEO films is studied using BET (Brunauer–Emmet–Teller) adsorption technique and reported as function of irradiation dose forS- and L-series with polymer concentration of 2 and 4 wt. %. A computer program [PROG$_{\text{IMAGE-POR}}$] is reported for determination of porosity and pore distribution of perturbed PEO films using SEM images and correlated with that obtained from BET technique. Novelty of PROG$_{\text{IMAGE-POR}}$ lies in the exposure of newer or undetected pore regime in which experimental pore-regime exists as a part. This method could be analyzed for mapping of porosity and average pore size.Practicability and repeatability of the developed programme has been established for monophasic system using backscattered imaging mode of SEM.Though, improvement this program is underdevelopment, but PROG$_{\text{IMAGE-POR}}$ bears the novelty of using image from either 2D-or 3D imaging system and could be applied for intricate composite/layered system.

**1.  Introduction**

Pores (or voids) are formed in polymer films irrespective of any preparation technique like gel casting, emulsion freeze drying, acylation, solution casting etc.[1, 2]. Easily soluble and fusible polymers are well used in the aforementioned techniques. However, certain conjugate polymer eg. polythiopene etc. being insoluble and infusible do not fit into such procedures. Conducting porous material is well studied due to their important applications in the field of electrolytic capacitors, battery separators and as electrode materials [3-5]. Furthermore, porous polymer films having micro-and sub micrometer pores have multiple applications in water purification, separation, scaffolds for tissue engineering, low dielectric constant materials for microelectronic devices, band gap materials, solid supports for sensors and catalysts etc. [6-12]. The basic challenge lies in the optimization of pore-size andporosity distribution based on the purpose of application. Preparation of two or three dimensional ordered porous structure is reported by Jiang et al. and Cassagneau et al. using templating method wherein self-assembled colloidal microspheres are utilized [13, 14].

In lieu of this above discussion, pore phase has recently emerged as an important subject to be well studied and optimized for multifaceted applications. Study of pore phase bearssignificance in identification of individual grain interfaces through the indirect outcome of the processes at micro-scale also termed as pore-scale by Wildenschildet al.[15]. Identification and investigation of such pore-scale is found to primarily govern the large scale phenomena in industrially viable materials eg. polymer, ceramic etc. Emphasis on the study of pore phase helps in,understanding the material properties, neglecting the outcomes at very high and smaller scales. Study of porosity in terms of quantification and distribution involved multiple well studied tools, viz. adsorption technique(BET technique, nitrogen adsorption)[16, 17], intrusion procedures as mercury porosimetry [18, 19], X-ray tomography etc. [20]. X-ray tomography has numerous advantages of generating 3-D information and for opaque porous media, about process and variables of importance to substrate flow and transport phenomena [21, 22]. In addition, tomographic imaging enables quantification of pore-scale [15]. In case of fluid flow and transport processes, technique's such as focused ion beam-scanning microscopy (FIB-SEM) and TEM tomography are in development stage wherein, representation of elementary volume is not well suited presently [15]. The major disadvantages associated with X-ray tomographic imaging includes artifacts produced by sample rotation and metal mountings, costly instrumentation, long measurement time for single sample, destructive nature for soft matter and large data files. In fact, real time imaging is not possible from tomographic technique, however, coded aperture imaging using a multi-slit code combined with CCD detectors provide real time imaging.

In the present article, the porous Poly(ethylene oxide), PEO films perturbed using high energy gamma dose (1 kGy -30 kGy, in powder and methanol solution state)is subjected to BET(Brunauer–Emmet–Teller)adsorption technique to study pore size distribution (PSD). Compared to numerous techniques employed for porosity and PSD viz. mercury porosimetry, Rutherford backscattering spectroscopy and small-angle neutron scattering (SANS) [23-25] etc., BET method is chosen due its simple and reliable measurement procedure [26].A novel computer program termed as **PROG$_{\text{IMAGE-POR}}$** is developed for determination of porosity, its size and distribution and is correlated with the experimental outcomes. The program, PROG$_{\text{IMAGE-POR}}$is based on the scanning electron microscopic (SEM) images as an input source file. Though there are certain limitations of using SEM 2D images, which fail to evolve all possible sample information, the approach is very simple and informative. The prime intention of the authors is not to compare the resultant porosity obtained from experimental BET and PROG$_{\text{IMAGE-POR}}$. However, the novelty of PROG$_{\text{IMAGE-POR}}$ lies in the exposure of newer or undetected pore regime. Experimental results on porosity and pore distribution have been standardized by several set of investigations fulfilling the criterion of reproducibility. Error analysis of these investigations has been carried out on the basis of standard deviation. The advantages as well as differences of the present program is discussed in details, however, the subsequent up gradation based on multiphasic systemis presently underway. Therefore, influence of high energy gamma irradiation on powder and solution of PEO is studied in terms of pore phase (% porosity and pore size distribution). Finally, effectiveness of the programis reported in terms of its application for intricate composites and effective in using whichever image from either 2D-or 3D imaging system.

## 2. Experimental

Self-standing polymer films (~ 200 $\mu$m) of Poly(ethylene oxide)(B.D.H., England, Mol Wt. $10^5$)were prepared using gamma irradiated powder and methanol solution of PEO. Samples in the form of PEO powder (termed as S-series) and methanol solution (L-series) were irradiated using $Co^{60}$ source with dose rate of 6.4 kGy.h$^{-1}$ in air. Gamma dose were varied in the range of 1-30 kGy followed by preserving the irradiated samples in air tight vacuum desiccator in order to protect from moisture and other external agents. The detail of film preparation by solution casting method using methanol as solvent is already described elsewhere [27] and sample identifications are given in Table 1, wherein the concentration of PEO was maintained at 0.02g.ml$^{-1}$ and 0.04 g.ml$^{-1}$ respectively.

For experimental determination of porosity and pore-size distribution (PSD), PEO films prepared from unirradiated; irradiated powder and methanol solution were subjected to BET analyzer [Quantachrome Instruments]. The microstructural images of the experimental polymer samples were characterized using scanning electron microscopy (SEM) [FEIC-QUO-35357-0614 with Bruker Quantax 100]. These SEM images were employed as an input source file for PROG$_{IMAGE-POR}$. The experiments were repeated on different part of single film to judge the standardization of film preparation. In the present context, the presented results are averaged from such multiple outcomes for all irradiation doses.In order to study the effectiveness and repeatability of the present programme, standard deviation of the average porosity are calculated for back scattered SEM images. Backscattered mode being the only parameter, dependent on atomic number of elements, standard deviation of average porosity is least and establishes the proposed programme to be universally accepted for monophasic system.

## 3. Details of the program (PROG$_{IMAGE-POR}$)

The basic flow chart for determination of porosity by PROG$_{IMAGE-POR}$ is shown in Figure 1. SEM images of experimental PEO samples (Table 1 in manuscript) are converted to *8bit type* image using ImageJ software and saved as *text image* with *.txt* file extension. Each of these text images contains integer numbers ranging from 0 to 255 as a 2D array with dimension *width\*height* of the SEM image. Integer numbers come from the pixel value of the corresponding *8bit* SEM image. Here 0 corresponds to perfect black colour and 255 corresponds to perfect white colour. Other numbers in between 0 and 255 gives the variation in gray scale. We have termed these numbers as gray scale pixel (GSP). We have set GSP to a particular threshold value GSP$_{th}$ to signify void (GSP >GSP$_{th}$) or solid (GSP <GSP$_{th}$) phase.

We have developed a **FORTRAN** program to find out the porosity value of the images setting GSP value where text images are the input files. In our simulation PROG$_{IMAGE-POR}$ GSP values are varied in the range of 50-127.5 (Figure 1). The outcome of PROG$_{IMAGE-POR}$ is obtained in the form of "*Pore size distribution [Porosity Intensity vs. Pixel dimension]*" and "*Average porosity [Av. Porosity vs. dose]*". The pixel dimension as obtained from simulation is converted to pore-dimension using Eq. 1 and 2. Equation 2 is calculated from IMAGEJ software.

$$Pore\ dimension\ in\ nm = Pixel\ Size * ({10^6}/{3.43}) * Magnification$$

(1)

$$where, 1\ Pixel = \left(\frac{1}{3.43}\right) * x\ in\ mm; \quad x = magnification \tag{2}$$

The SEM images being two dimensional, pose a problem in the selection of depth of the pore phases as it lacks the third dimension. We can get better result by stacking sequentially image of each layer of the polymer sample if possible.

## 4. Results and Discussion

### 4.1. Porosity and Pore-size distribution of PEO films obtained from γ-irradiated powder and methanol solution: Study using Adsorption Technqiue

The distribution of pore phase in perturbed pristine PEO system is found to depend on the physical state subjected to gamma irradiation. The present section illustrates the influence of gamma irradiation on porosity and pore size distribution of PEO films studied using BET technique. The average dimension of pores lie within 100-500 nm for PEO films prepared using irradiated powder (2/4-S series) and/or methanol solution (2/4-L series) as observed from Figures 2 and 3. In addition to the mentioned pore-dimension regime, smaller pores of <100 nm exists for unirradiated PEO matrix which disappear upon subsequent irradiation. The extent of multimodal pore distribution is high for 2S- and 4S-series at lower irradiation doses which subsequently becomes unimodal with much smaller pores with increment in dose. Similar trend is observed for 2L- and 4L-series (Figure 3).Increase in irradiation dose generates excited radicals which stabilize through mobilization of polymer chains thereby forming larger spherullites. Growth in the size of spherullites minimizes pore phase (low % porosity) and shifts pore dimension to lower regime (from right to left in Figures 2-3) with increase in irradiation. In the present context, air assisted irradiation of either PEO powder or methanol solution promotes scission as already established in our earlier communication [27]. However, most of the gamma radiation is absorbed by the solvent during solution irradiation which generates –CH$_2$OH radical that acts as a cross linking agent owing their higher mobility. Consequently, % porosity is lower for both 2/4 L-series compared to 2/4 S-series (Figure 4).

Irrespective of powder or solution state irradiation, predominant contribution of scission increases % porosity sharply till 5kGy in all the samples. However, with further increase in dose, matrix densification results due to the formation of innumerable scission fragments in S-series. 2/4L-series possess much lower porosity compared to S-series with sharp declining trend after 5 kGy. Compared to powder irradiation, high energy perturbation of PEO solution enables a pattern in pore distribution having lesser modality (uniform distribution). As expected, more concentrated 4L-series is more denser (low porosity) with sharp declining porosity after 5 kGy. The significant outcome as observed from Figure. 2-4 is that, selective regime of % porosity with unique pore-size distribution is obtained with a specific irradiation dose and state of polymer. This helps in selection of experimental polymer system for explicit application.

## 4.2. Implementation of PROG$_{\text{IMAGE-POR}}$ in Pristine Perturbed PEO

Experimental tools are though specific; possess certain limitations in estimation of porosity of soft polymer films. The instrumentation for BET technique is based on the inert gas adsorption through the pores of PEO film followed by fitting to certain standard isotherm. The outcome in the form of pore size distribution is therefore indirectly dependent on the type of isotherm which closely matches with the experimental results. Conceptually, such experimental techniques might either rule out or encompass certain non-existent pore sizes within the sample. The present section intend to implement a computer program [PROG$_{\text{IMAGE-POR}}$] based on two dimensional scanning electron microscopy (SEM) images of PEO films synthesized with irradiated power and/or methanol solution. The program is executed on five selective magnifications from 100x to 10000x for each experimental polymer films. Based on the grey scale image of SEM, ImageJ software converts it into *.txt* format. After that the program considers variable regime of grey scale pixel (GSP) ranging from 60 to 127.5 with 0 for pure black and 255 for pure white.Pores are considered to be *white*and matter (polymer) to be *black*. In such variable GSP, the program is used to study the distribution of *WHITE* (pore phase) within the SEM.*txt* files. As magnification of scanning electron microscopy enhances from 100x to 10000x, newer facts/informationregarding the image [bulk or pore]are revealed and prior evidences get wiped off.Optimum threshold of GSP is considered to be 70 upon comparison with the % porosity studied from BET technique as shown in Figure 5. The average porosity obtained from GSP 60 is too low, whereas higher limit of GSP 80 and 127.5 shows much higher porosity with erratic distribution which is irrelevant with the magnitude and distribution of the PEO film porosity. The importance of PROG$_{IMAGE-POR}$ is in unrevealing the pore-dimensions as a function of SEM magnification for the irradiated samples rather than comparing the magnitude and distribution of pores with that obtained from BET. One of the major shortcomings of PROG$_{IMAGE-POR}$ is in the: a) consideration of 2D images for porosity calculation and b)using grey scale image having pixel value within 0-255 rather than binary image. However, the prime intention of the authors is to discuss the initial simplest approach of PROG$_{IMAGE-POR}$ which is capable of understanding the influence of gamma dose on the pore phase of PEO matrix using any specific image obtained by any technique as revealed in Figure 6. The useful imaging modes as shown in Figure 6 enable easy determination of sample porosity, but possess certain limitations which restrict the general applicability. The reported program possess the flexibility in considering image obtained of any one of the 3d tool described in Figure6 containing specialized sample information. Hence, all these special evidences obtained from selective 3D technique could be generalized using PROG$_{IMAGE-POR}$ for effective determination of porosity or any other feature of the sample. The present endeavour is generalized in which further improvements in imposing such realistic assumptions are underway so that entire matrix information could be easily revealed.

Figures 7-10 describes the pore-size distribution obtained from PROG$_{\text{IMAGE-POR}}$ using SEM images for 2/4-S and 2/4-L series for 1, 10 and 30 kGy in five magnification regimes from 100x to 10000x respectively. The representative SEM images for the particular dose are also given in the inset at only one magnification. The given SEM images are only representative of the polymer sample at a particular irradiation dose. During programming, multiple images from identical magnification are given as input in order to enhance the accuracy. As expected, irrespective of sample type, it could be visualized that the size distributions obtained from PROG$_{\text{IMAGE-POR}}$ (Figures 7-10) are not exactly similar with the experimental distribution pattern (Figures 2-3). With increase in magnification for a particular dose, smaller pores ($10^2$-$10^3$ nm) are revealed. It could be noted that experimental BET techniques shows pore size within 100-500nm i.e $10^2$ nm. However, this experimental regime of pore dimension exists only at higher magnification as calculated from PROG$_{\text{IMAGE-POR}}$. At still lower magnification from 100x to 5000x, larger pores within $10^3$-$10^5$nm (0.1 mm) are found to exist. It could be noted that, the intensity of larger pores (~$10^4$-$10^5$ nm) is higher for 30 kGy dose irrespective of sample type compared to the lower irradiation doses. This is in agreement to the predominance of scission at higher gamma dose exposed in air atmosphere. The comparison of pore distribution at 5000x for 2/4S- and 2/4-L series is shown in Figure 11 as a function of gamma dose of 1kGy (Figure 11a), 10 kGy (Figure 11b) and 30 kGy (Figure11c) respectively. This graph elucidates the influence of gamma dose on the nature of pore distribution of all samples at intermediate magnification of 5000x.It is observed that 50000 x magnification unveils pore in the dimension of ~ $10^3$ nm irrespective of sample type. Lower dose of 1 kGy generates uniform distribution of pore with intermediate intensity. Population (i,e intensity) of pores is found to increase with dose

increment to 10 kGy specially for 2/4S-series. Erratic distributions are obtained with highest dose of 30 kGy which generates smaller pore (less than $10^2$ nm) for 2S-series and larger pores ($> 10^3$ nm) for 2L-series. It has already stated previously that aim of the present program based on the image of sample is to unveil the newer pore dimension existing within the sample and not to compare the magnitude of porosity trends with the experimental findings. Such discernment among experimental BET technique and the results obtained from PROG$_{\text{IMAGE-POR}}$ is shown in Figure 12 c1-c4. Figure 12a and b shows average porosity (%) obtained from experimental BET study and that from PROG$_{\text{IMAGE-POR}}$ considering optimum GSP of 70. The magnitude shown in the pie charts in Figures 12 c1-c4 does not indicate error regime, instead clarify that the entire pore dimension in irradiated polymer films is not revealed in the experimental which account for the shown differences.The practical applicability of such programme is further established using simulation of backscattered SEM images for PEO system.

### 4.3.Feasibility of PROG$_{\text{IMAGE-POR}}$for any monophasic system

Detailed discussion on the application of **PROG$_{\text{IMAGE-POR}}$ for** irradiated PEO system (in solid and liquid phase) in Figures 5-12 enables the study of porosity and pore size distribution in the regimes which are not revealed through BET technique. However, reliability and repeatability is an issue for such SEM based methods which imbibes variation in contrast, brightness, user variability and instrument dependence. Owing to such fact, we have taken SEM micrographs for the experimental samples using two modes:

a) Secondary mode (SE) represented by e =0

b) Back scattered mode (BSE) represented by e =1

BSE images are limited to a grayscale range because they only record one variable, average Z (atomic number,signature of element variation). Therefore, "brighter" BSE intensity correlates with greater average Z in the sample, and "dark" areas have lower average Z. BSE images are very helpful for obtaining high-resolution compositional maps of a sample and for quickly distinguishing different phases. In SE modes, electron beams are scattered from the surface of sample. However, during BSE mode, the scattering of electron beam results from the volume element of the sample and thereby consists of information about the three dimensional morphology as a function of atomic number. Consequently, In light of such background, at a fixed contrast, brightness of the BSC modes has been altered as per Table 2. Average porosity and the corresponding standard deviation is determined using variable grey scale threshold ($GSP_{th}$) ranging from 60 to 120 and are shown in Table 3.The respective SEM images based on the mentioned three conditions of BSE mode are shown in Table 2 along with the variation of $P_{av}$ *[including Sd]* **with** GSP is shown in Figure 13.

In comparison to the results of BSE mode, simulation results are also shown for secondary mode of imaging (e = 0). The details of parameters for SE mode and variation of average porosity as a function of GSP is shown in Table 2, 4and the micrographs are shown in Figure 14.It is to be noted that, the magnitude of average porosity varies within a narrow range with a reducing error (Sd) with $GSP_{th}$ increment for BSE mode. The magnitude of average porosity is satisfactory for GSP $_{th}$ below 90 for BSE. However, the present program is found is be more relevant upon incorporation of backscattered mode in SEM imaging as seen from the mentioned results. These results are reproducible in terms of both person, SEM condition and machine variability.

Therefore, the reliability of the porosity and pore size distribution as determined from the computer program [PROG$_{IMAGE-POR}$] is established and can be well applied for mono phasic systems. In case of multiphase, parameters need to be re verified and tested accordingly.

### 4.4.Importance of PROG$_{IMAGE-POR}$ for Intricate Matrices

The aforementioned section elucidates application of PROG$_{IMAGE-POR}$ on porosity and its distribution of PEO films prepared with gamma irradiated powder and methanol solution.

Being a film with uniform morphology throughout, the study of porosity and its distribution using any technique is fairly simple. However, in experimental material science, complex structures do exits whose study in terms of porosity is difficult. An account of such matrices is shown in Figure 14. Variety of thin films may be formed on different substrates, which show intricate layered structures. In this context it is noteworthy to mention graphene-molybdenum di sulphide based films having such intricate layered structures whichare widely used in electrochemistry [28]. In these structures specially Figure14b and c, morphology and porosity of individual layer is different and in combination form a functional material. Study of porosity for such material is stringent since experimental tools give an average porosity of the monolith. In other way, each layer has to be differently prepared and tested, wherein the properties of total monolith on individual layer porosity gets shaded. In such situation, the present PROG$_{IMAGE-POR}$ is expected to be functional with different inputs of images from variable imaging tools in different orientations. Work in this direction involving intricate morphology and regarding up gradation of the program is under process.

## 5. Conclusion

A new, simple and novel computer program termed as PROG$_{IMAGE-POR}$ is reported for study of porosity and its distribution of Poly(ethylene oxide) (PEO) films prepared with gamma irradiated [1-30 kGy] powder (2/4 S-series) and methanol solution (2/4L-series). The algorithm uses any sort of sample image as the input source file. The morphology could be obtained from any 2D-or 3D imaging software e.g. SEM, TEM, tomographic image etc. In the present case SEM micrographs of perturbed polymer films are used as the source file. It is found that the experimental data related to porous structure of PEO films obtained from BET (100-500 nm)measurements appears to be a part of the entire range of pore spectrum (~ $10^2$-$10^5$ nm) obtained from 2D SEM image based program, PROG$_{IMAGE-POR}$. With increase in the SEM magnification, newer information regarding porosity is revealed and prior observations get wiped off. The study on porosity of the polymer films are thus, studied as a function of gamma irradiation dose, irradiation state, concentration of polymer and magnification of imaging tool (SEM) from PROG$_{IMAGE-POR}$. Feasibility of the present computer programme is further established using standard deviations simulated from back scattered SEM (BSE) images for monophasic PEO system. BSE images being dependent on the atomic number of elements, provides relevant information about the image. Standard deviation in average porosity is obtained upon simulation of BSE images which confirms the applicability of programme for monophasic system.Finally, applicability of this simple program is reveled in case of certain intricate microstructures of thin films, layered structure etc. wherein the usual techniques are restricted. However, significant upgradation of this program is in progress so as to enhance its applicability.

**Acknowledgements**

SKP is thankful to Inter-University Accelerator Centre (IUAC), Delhi for providing Junior research fellowship. Dr. Paramita Bhattacharyya and Miss. Suchisrawa Ghosh of Department of Food Technology, Jadavpur University, Kolkata are acknowledged for extending their support in using the gamma irradiation chamber.The Authors also acknowledge FIST-2, DST Government of India, at the Physics Department, Jadavpur University for providing the facility of SEM microscope.

# Tables:

**Table 1:** Sample identification of experimental Poly(ethylene oxide) films

| Sample Irradiated | Concentration of PEO (g.ml⁻¹) | Sample ID | | | | | |
|---|---|---|---|---|---|---|---|
| | | Dose (kGy) | | | | | |
| | | **1** | **5** | **10** | **15** | **20** | **30** |
| Powder | 0.02 | 2S-1 | 2S-5 | 2S-10 | 2S-15 | 2S-20 | 2S-30 |
| Irradiation | 0.04 | 4S-1 | 4S-5 | 4S-10 | 4S-15 | 4S-20 | 4S-30 |
| Solution | 0.02 | 2L-1 | 2L-5 | 2L-10 | 2L-15 | 2L-20 | 2L-30 |
| Irradiation | 0.04 | 4L-1 | 4L-5 | 4L-10 | 4L-15 | 4L-20 | 4L-30 |

**Table 2:** Details of parameters for various SEM imaging modes

| | **Contrast** | **Brightness** |
|---|---|---|
| BSE mode [ e=1] | 60 | 50 |
| | 60 | 60 |
| | 60 | 75 |
| | 55 | 60 |
| SE mode [e = 0] | 65 | 65 |
| | 70 | 75 |

**Table 3:** Average porosity as a function of GSP for BSE mode [e=1]

| GSP threshold | Average Porosity [$P_{av}$] | Standard deviation (Sd) |
| --- | --- | --- |
| 60 | 0.374 | 0.096 |
| 70 | 0.382 | 0.085 |
| 80 | 0.392 | 0.074 |
| 90 | 0.403 | 0.062 |
| 100 | 0.414 | 0.048 |
| 110 | 0.427 | 0.033 |
| 120 | 0.441 | 0.017 |

**Table 4:** Average porosity as a function of GSP for SE mode [e=0]

| GSP threshold | Average Porosity $P_{av}$ | Standard deviation (Sd) |
| --- | --- | --- |
| 60 | 0.325 | 0.043 |
| 70 | 0.368 | 0.075 |
| 80 | 0.412 | 0.154 |
| 90 | 0.452 | 0.232 |
| 100 | 0.488 | 0.298 |
| 110 | 0.517 | 0.348 |
| 120 | 0.514 | 0.389 |

# Figure Captions:

**Figure 1.** Schematic flow chart for the developed program [PROG$_{\text{IMAGE-POR}}$]

**Figure 2.** Pore-size distribution determined from BET technique for: a) 2S- and b) 4S-series as a function of irradiation dose.

**Figure 3.** Pore-size distribution determined from BET technique for: a) 2L-and b) 4L-series as a function of irradiation dose.

**Figure 4.** Average porosity (in %) as a function of irradiation dose for 2/4S- and 2/4L-series obtained from BET technique.

**Figure 5.** Average porosity (in %) as a function of irradiation dose for 2/4S- and 2/4L-series obtained from PROG$_{\text{IMAGE-POR}}$ for grey scale threshold (GSP$_{\text{th}}$): a1) 60, a2) 70, a3) 80 and a4) 127.5. Close matching of a2) GSP$_{\text{th}}$ 70 is shown with b) average porosity of 2/4S- and 2/4L-series as determined by BET

**Figure 6.** Schematic representation for list of 3D-imaging modes with applicability of the program PROG$_{\text{IMAGE-POR.}}$

**Figure 7.** Pore-size distribution obtained from PROG$_{\text{IMAGE-POR}}$ as a function of SEM magnification for 2S-series irradiated at: a) 1kGy, b) 10 kGy and c) 30 kGy. *The inset figures corresponds to the SEM micrographs of the concern irradiation dose at 2500x magnification*

**Figure 8.** Pore-size distribution obtained from PROG$_{\text{IMAGE-POR}}$ as a function of SEM magnification for 2L-series irradiated at: a) 1kGy, b) 10 kGy and c) 30 kGy. *The inset figures corresponds to the SEM micrographs of the concern irradiation dose at 2500x magnification.*

**Figure 9.** Pore-size distribution obtained from PROG$_{\text{IMAGE-POR}}$ as a function of SEM magnification for 4S-series irradiated at: a) 1kGy, b) 10 kGy and c) 30 kGy. *The inset*

*figures corresponds to the SEM micrographs of the concern irradiation dose at 2500x magnification*

**Figure 10.** Pore-size distribution obtained from PROG$_{\text{IMAGE-POR}}$ as a function of SEM magnification for 4L-series irradiated at: a) 1kGy, b) 10 kGy and c) 30 kGy. *The inset figures corresponds to the SEM micrographs of the concern irradiation dose at 5000x magnification*

**Figure 11.** Pore-size distribution obtained from PROG$_{\text{IMAGE-POR}}$ for 2/4S- and 2/4L-series at 5000x magnification irradiated at:a) 1kGy, b) 10 kGy and c) 30 kGy.

**Figure 12.** a1)-a4) Analysis of variation in average porosity among experimental BET and PROG$_{\text{IMAGE-POR}}$ as a function of irradiation dose for: c1) 2S-, c2) 4S-, c3) 2L- and c4) 4L-series.

**Figure 13.** SEM images of 4S-series_15 kGy in BSE mode for1500 x magnification using: a) C= 60; B =50, b) C= 60; B =60, c) C= 60; B =75 and d) variation of average porosity (with Sd) as a function of GSP.

C= Contrast, B= Brightness and Sd = standard deviation.

**Figure 14.** SEM images of 4S-series_15 kGy in SE mode for1500 x magnification using: a) C= 55; B =60, b) C= 60; B =65, c) C= 70; B =75 and d) variation of average porosity (with Sd) as a function of GSP.

C= Contrast, B= Brightness and Sd = standard deviation.

**Figure 15.** Schematic of some complex matrixes viz. a) thin film categories, b) layered structure and c) Heterostructural bilayers onto which PROG$_{\text{IMAGE-POR}}$ could be proved functional

**Figure 1**

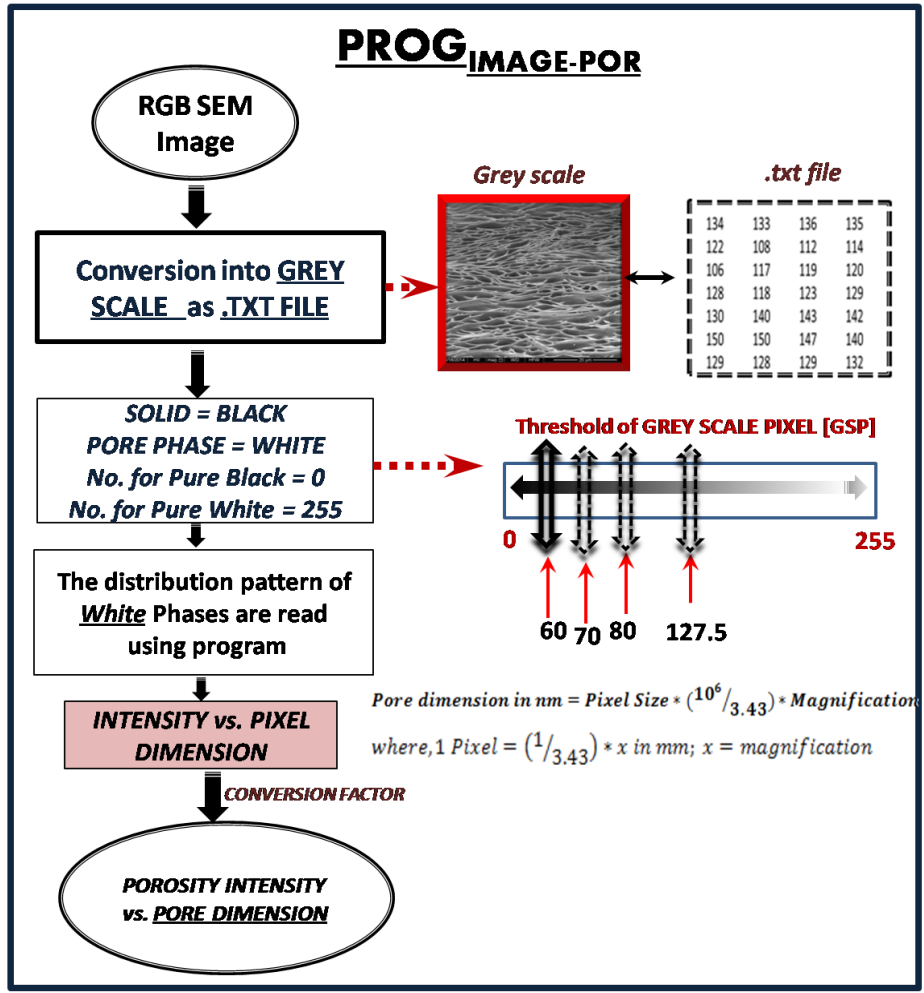

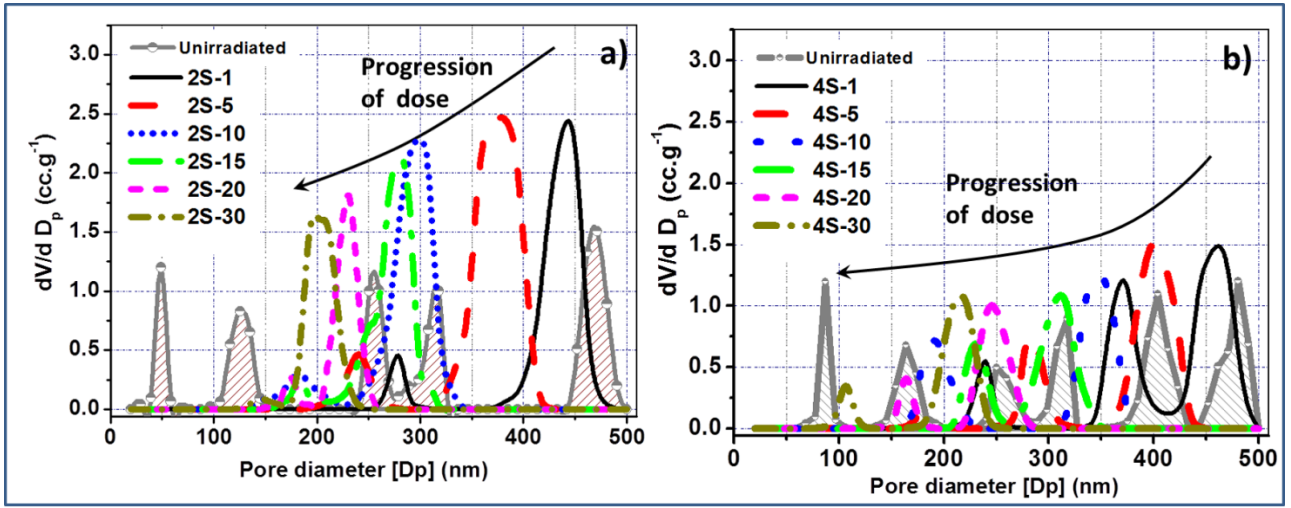

**Figure 2**

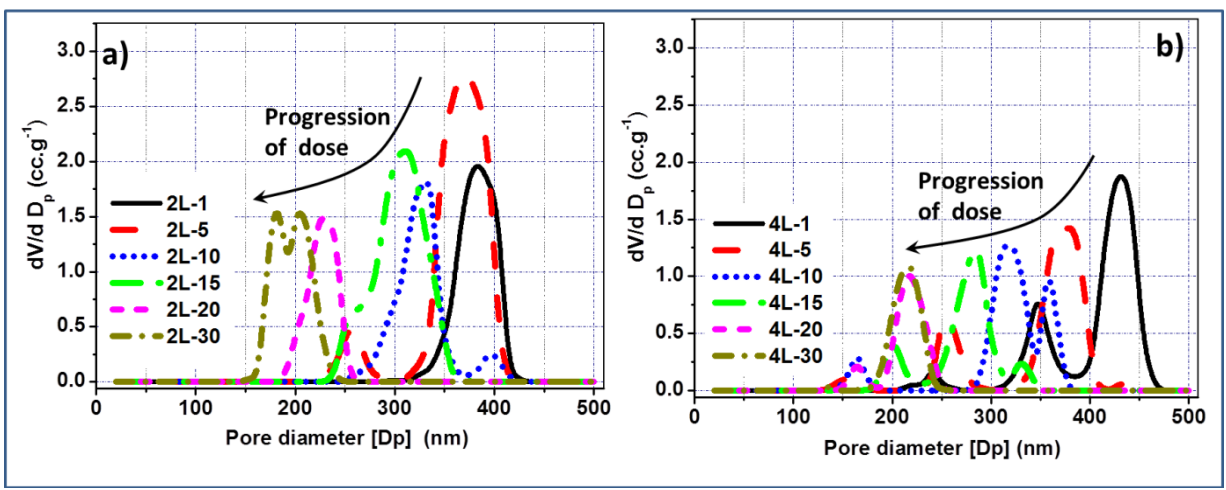

**Figure 3**

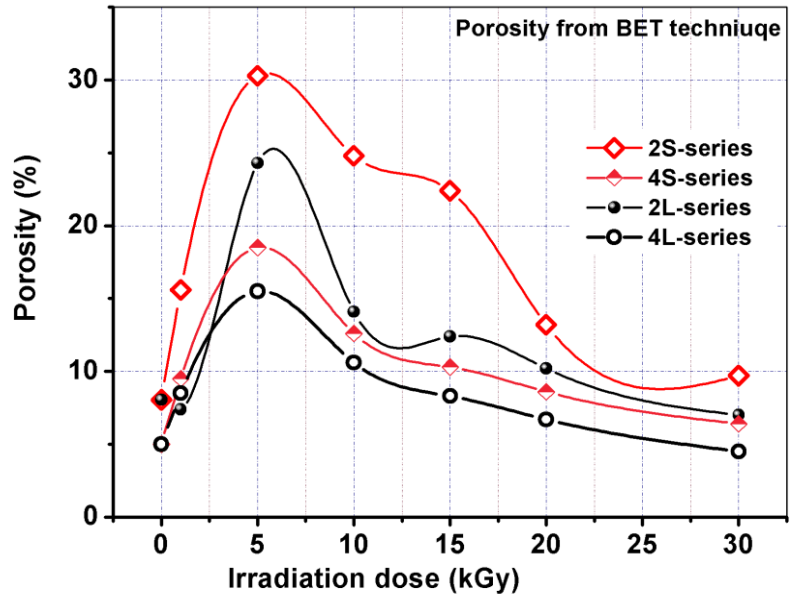

**Figure 4**

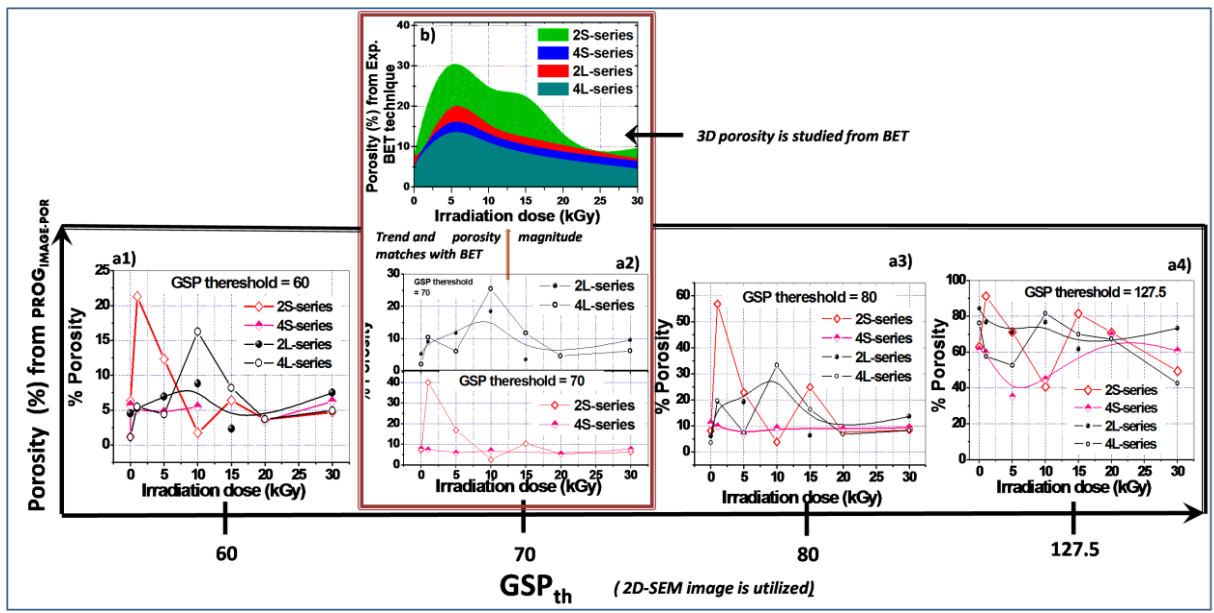

**Figure 5**

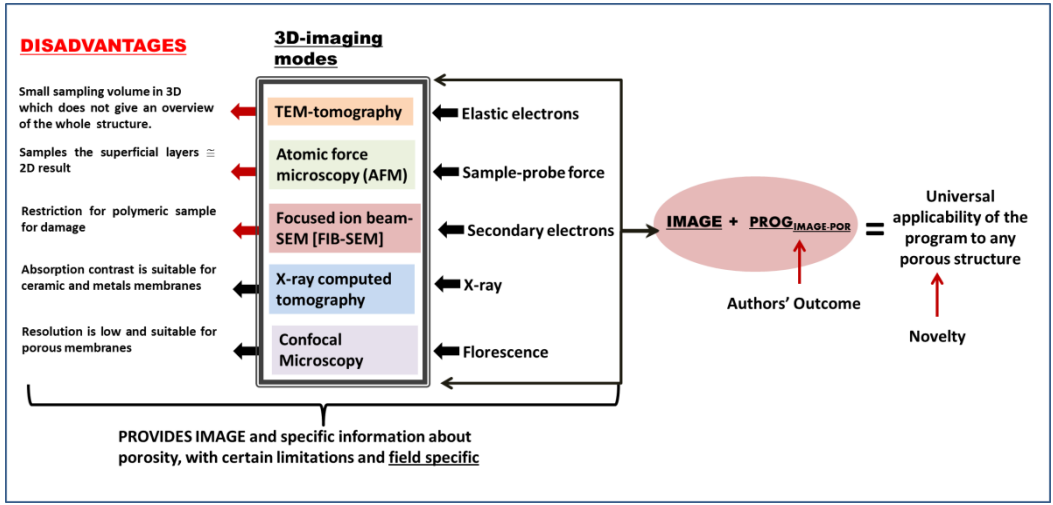

**Figure 6**

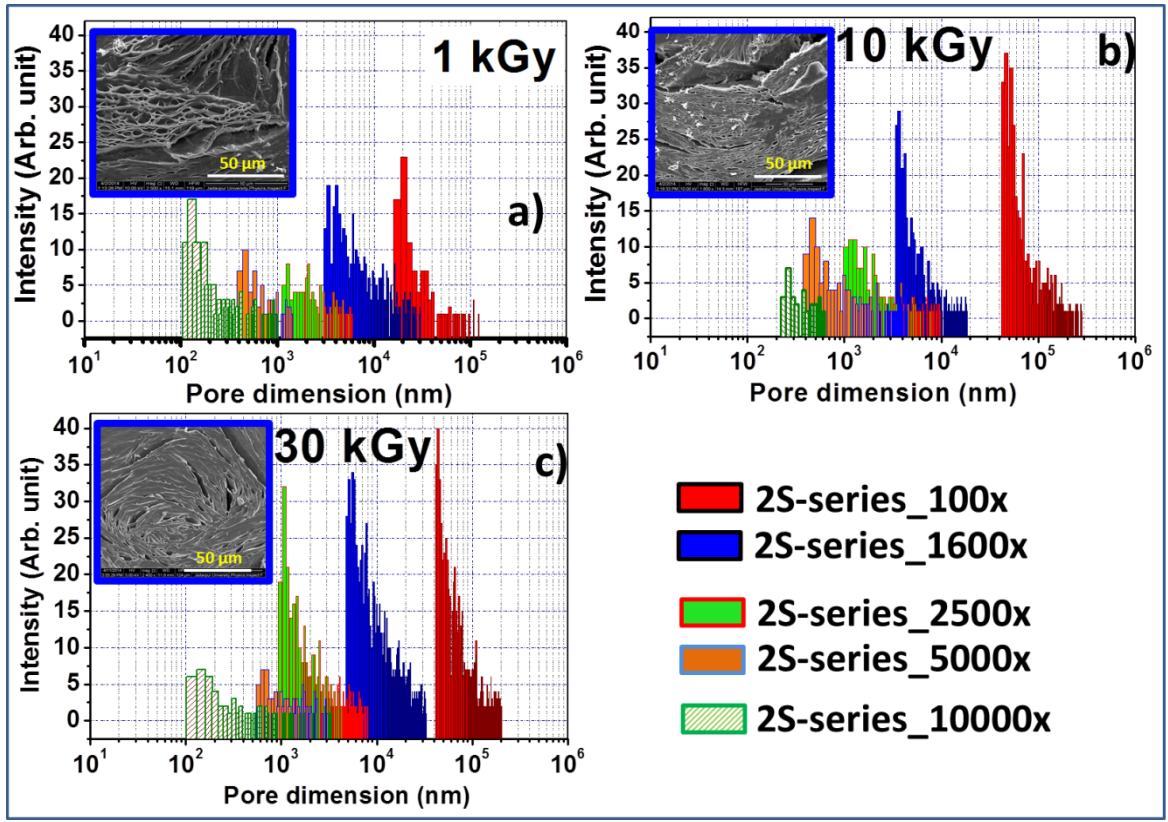

**Figure 7**

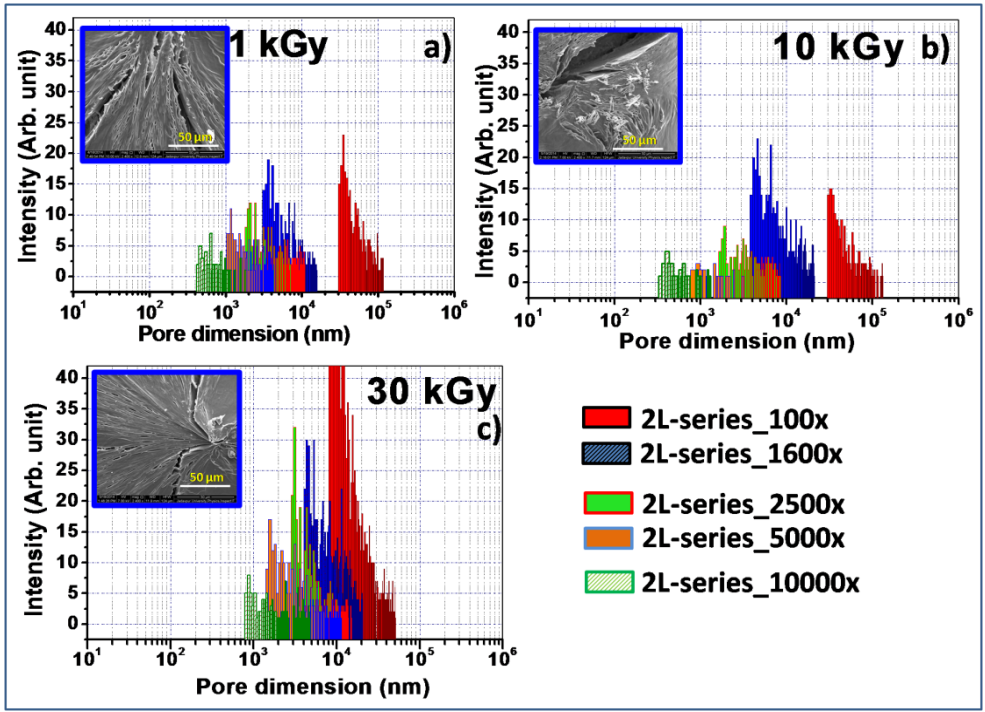

**Figure 8**

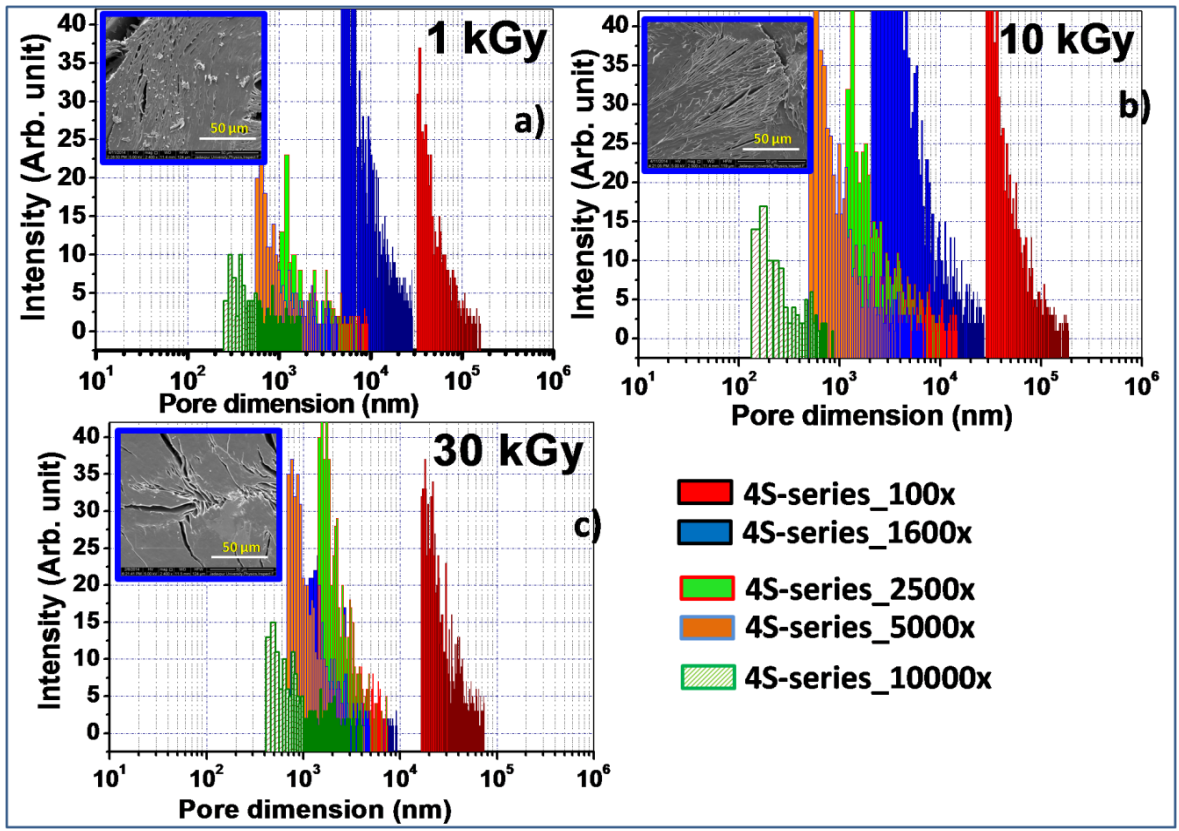

**Figure 9**

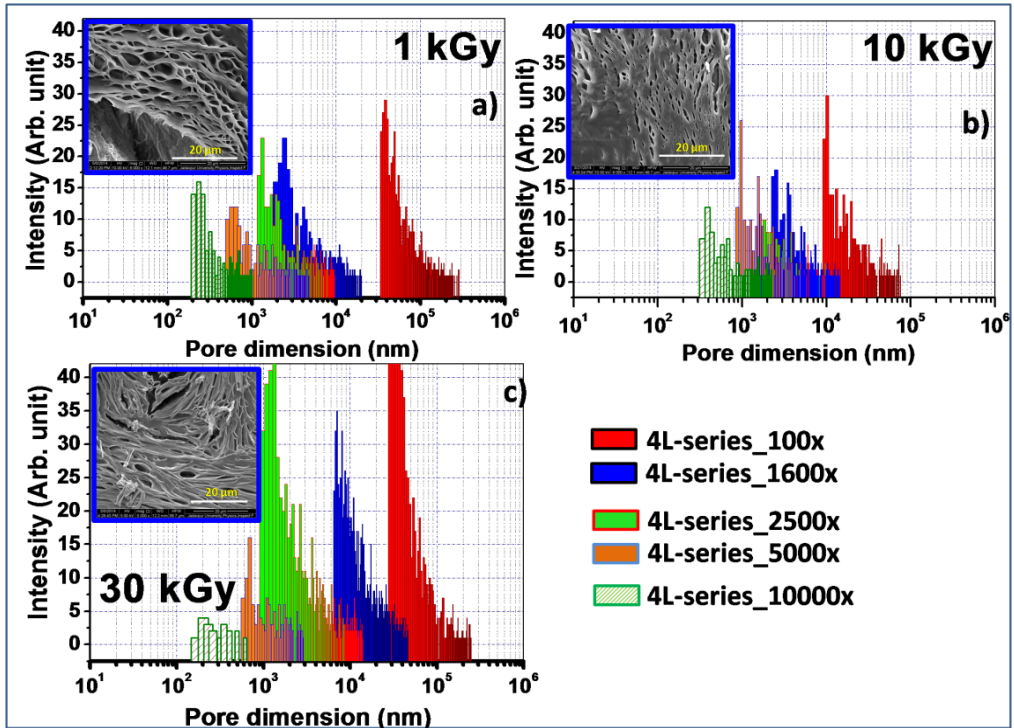

**Figure 10**

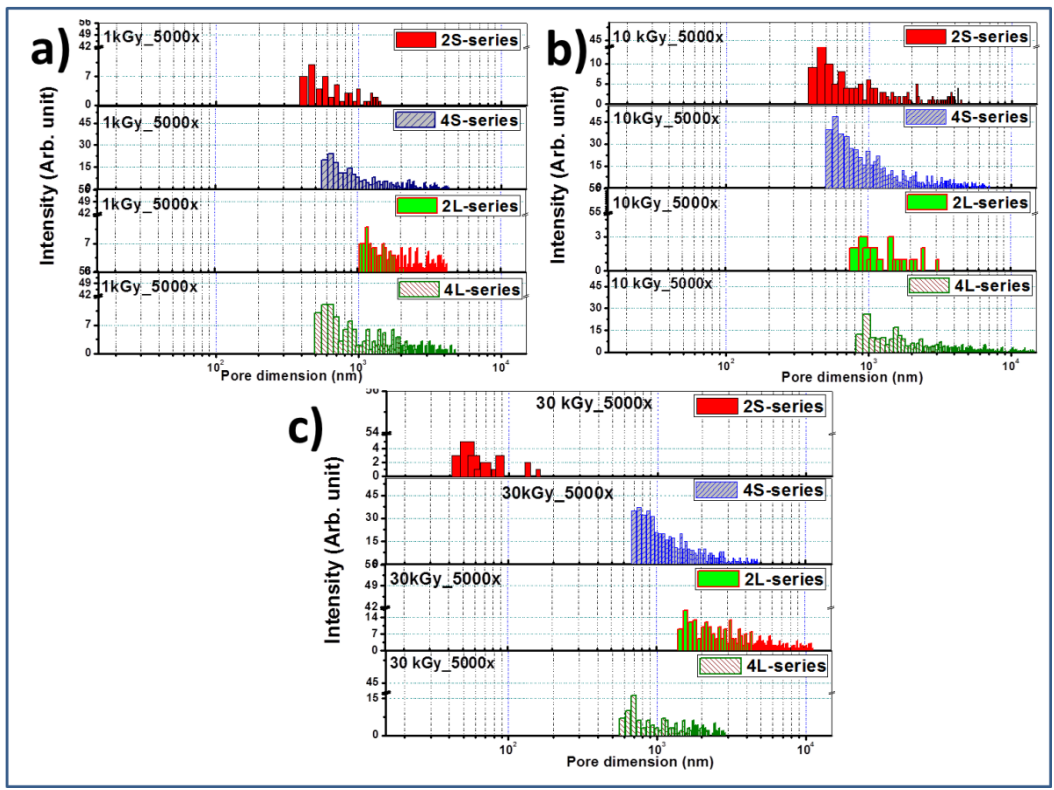

**Figure 11**

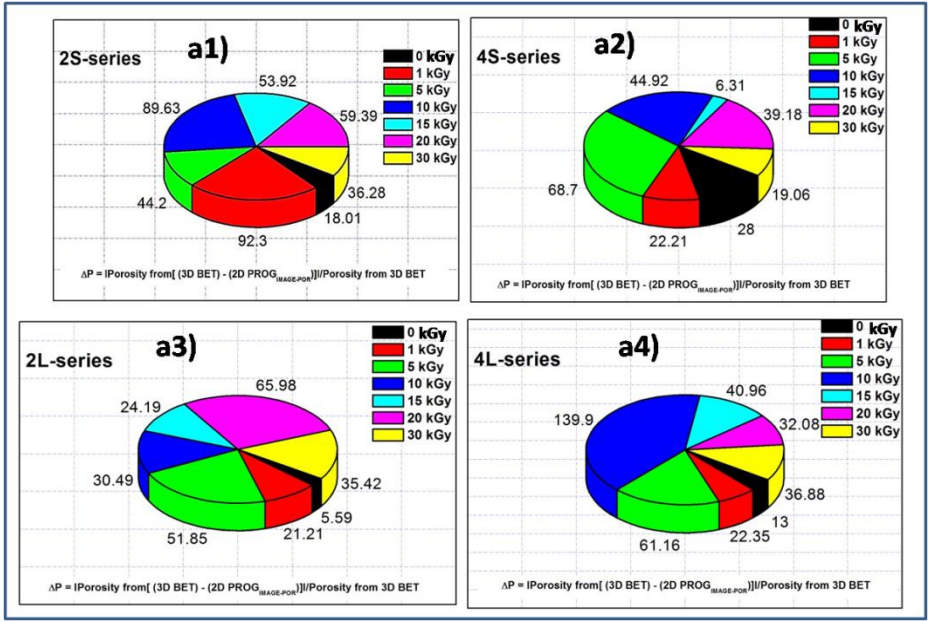

**Figure 12**

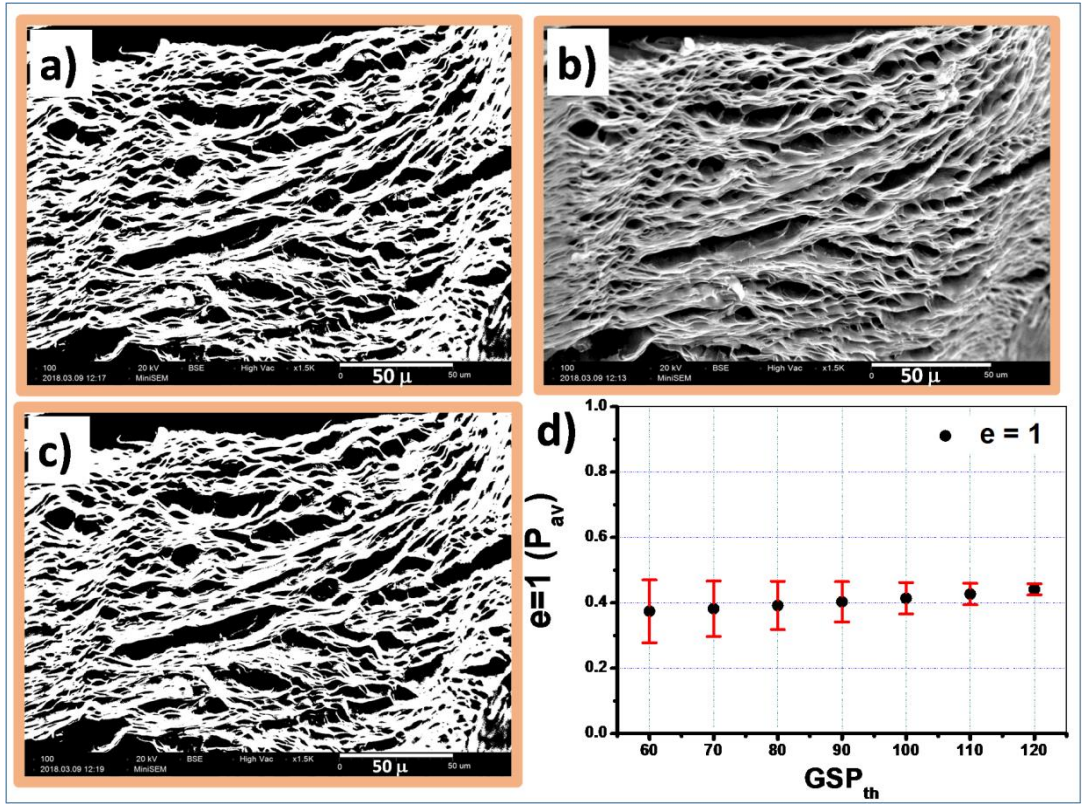

**Figure 13**

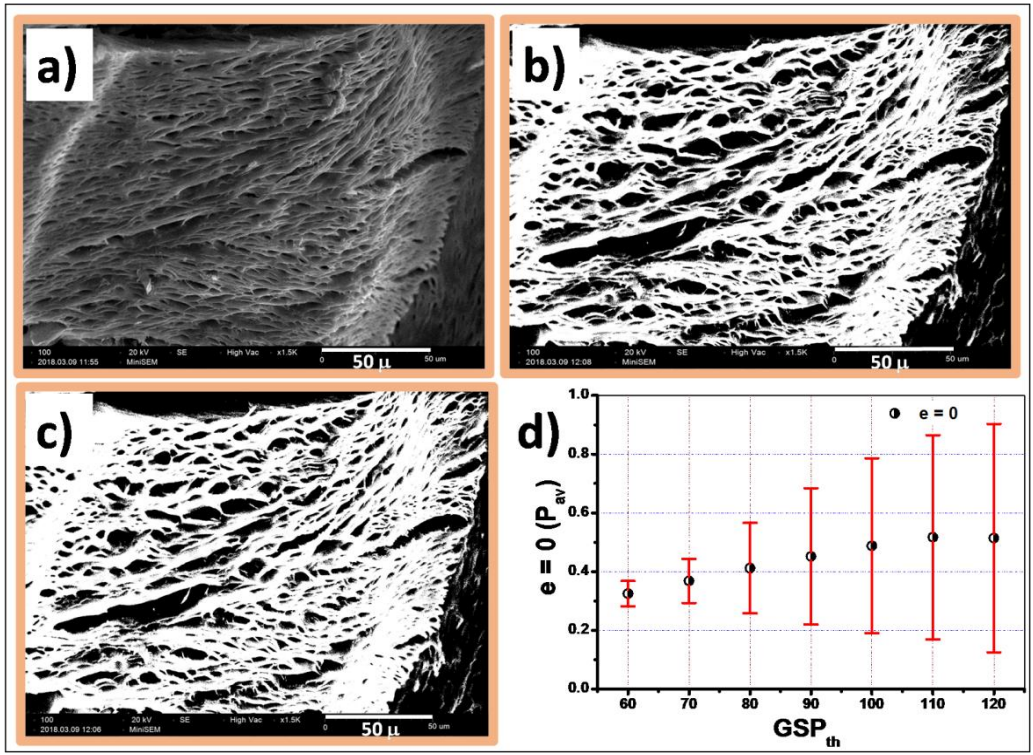

**Figure 14**

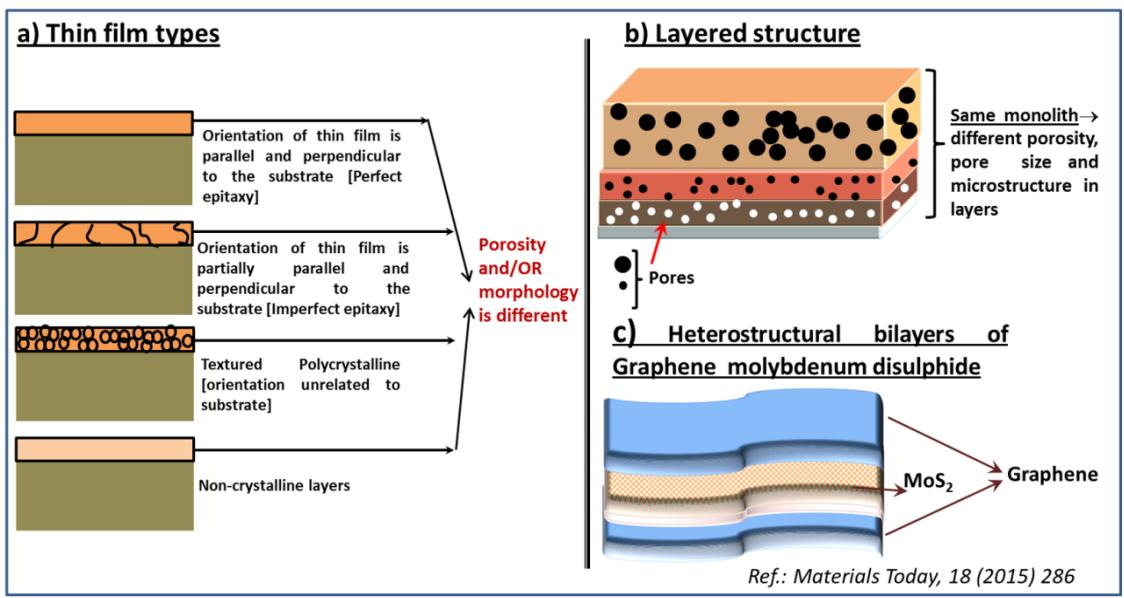

**Figure 15**